\documentclass[a4,12pt]{article}
 
\usepackage{amsmath,amssymb}
\usepackage{cite}
\usepackage{color}

\definecolor{color1}{rgb}{0,0,0.5}
\usepackage{hyperref}
\hypersetup{colorlinks,
            linkcolor=color1,
	    citecolor=color1,
	    urlcolor=color1
	   }

\newcommand{\Tr}{{\rm Tr\,}} 
\renewcommand{\d}{{\rm d}} 
\renewcommand{\i}{{\rm i}} 
\newcommand{\e}{{\rm e}} 

\title{\bf Generalized BF state in quantum gravity}
\author{
Shinji Yamashita\footnote{
Email: {\tt s\_yamashita@kumadai.jp}
}\ , 
Satoshi Yajima, and Makoto Fukuda \\
{\small \it Department of Physics, 
Kumamoto University, Kumamoto 860-8555, Japan}
}
\date{}

\begin{document}
\maketitle
\begin{abstract}
 The BF state is known as a simple wave function that satisfies 
 three constraints in canonical quantum gravity without 
 a cosmological constant.
 It is constructed from a product of the group delta functions.
 Applying the chiral asymmetric extension,
 the BF state is generalized to the state for real values
 of the Barbero--Immirzi parameter.
\end{abstract}

\section{Introduction}
In modern canonical quantum gravity,
the connections with the Barbero--Immirzi parameter $\beta$ play
the role of fundamental variables
\cite{Barbero1995, Immirzi1997, Ashtekar2004}.
Wave functions are 
required to solve the three constraints, i.e., Gauss, diffeomorphism, 
and Hamiltonian constraints.

The Chern--Simons (CS) state,
which is also called the Kodama state, is known as an exact solution 
of these three constraints with a cosmological constant 
\cite{Kodama1990}.
In this case, the configuration variable is a complex 
$sl(2, {\mathbb C})$-valued connection, which takes a left- or 
right-handed form, namely, $\beta=\pm \i$ for the Lorentzian case.
However, loop quantum gravity (LQG) proposes that
the Barbero--Immirzi parameter takes real values for several
technical reasons.
The real value of $\beta$ gives a real $su(2)$-valued connection,
but it makes the Hamiltonian constraint more complicated.

The generalization of the CS state to real values of $\beta$ 
was achieved by Randono \cite{Randono2005, Randono2006, Randono2006a}.
The generalized states are parameterized by the Levi-Civita curvature,
and solve some difficulties of the ordinary CS state,
e.g., the normalizability and the reality conditions.

On the other hand, 
the wave function without a cosmological constant was found by 
Mikovi\'c \cite{Mikovic2003, Mikovic2004}.
It is called the BF state here.
This state is given as a product of the group delta functions of the
curvature, and constructed from the left-handed complex 
connection as well as the ordinary CS state.
In this paper, the generalization of the BF state is considered
as an analog of the generalization of the CS state.
We would like to emphasize that the process of the generalization 
follows Refs. \cite{Randono2005, Randono2006, Randono2006a}.
Specifically, it is carried out via the chiral asymmetric extension.

In Sect. 2, we briefly review the BF state for $\beta = \i$.
Then, using the chiral asymmetric model,
the BF state is extended to the case of generic purely imaginary 
values of $\beta$.
In Sect. 3, the BF state is extended further to the case of generic 
real values of $\beta$.
This state is expressed in terms of the real $su(2)$-valued 
connection and the Levi-Civita curvature. 
Making use of the appropriate inner product,
three constraints with real values of $\beta$ are solved.
In Sect. 4, we present the conclusions and discuss the results. 

\section{BF state and chiral asymmetric extension} 
\subsection{BF state}
The three constraints of canonical quantum gravity without 
a cosmological constant can be derived from the Holst action 
\cite{Holst1996}
\begin{equation}
 S_{\rm H} 
  = \frac{1}{4k}\int 
  \left(
   \epsilon_{IJKL}\, e^{I} \wedge e^{J} \wedge \Omega^{KL}
   - \frac{2}{\beta}\, e^{I} \wedge e^{J} \wedge \Omega_{IJ} 
  \right) \ , 
  \label{eq:HolstAction}
\end{equation}
where $k=8\pi G$, $e^{I}$ is the tetrad, and 
$\Omega = \d \omega + \omega \wedge \omega$ is the curvature of 
the spin connection $\omega^{IJ}$.
Capital Latin indices $I, J, \dots$ are used as Lorentz indices.
Performing the Legendre transformation, one obtains
\begin{equation}
 S_{\rm H}
  = \frac{1}{k\beta}\int \d^{4}x
  \left(
   E_{i}^{a}\dot{A}^{(\beta)}{}_{a}^{i} 
   + \lambda^{i}G_{i} + N^{b}V_{b} + NH
  \right) \ ,
\end{equation}
where 
$\dot{A}^{(\beta)}{}_{a}^{i} 
= {\cal L}_{t}A^{(\beta)}{}_{a}^{i}$\ ,
$\lambda^{i}, N^{a}$, and $N$ are Lagrange multipliers, 
and $G_{i}, V_{b}$, and $H$ are Gauss, diffeomorphism, 
and Hamiltonian constraints respectively.
Letters $i, j, \dots$ and $a, b, \dots$ denote 3D 
internal and spatial indices, respectively.
The configuration variable 
$A^{(\beta)}{}_{a}^{i} = \Gamma_{a}^{i} + \beta K_{a}^{i}$ 
is constructed from the Levi-Civita spin connection $\Gamma_{a}^{i}$ 
and the extrinsic curvature $K_{a}^{i}$. 
Choosing the time gauge $e_{a}^{I}|_{I=0}=0$, 
the canonical momentum variable can be written as 
$E_{i}^{a}=\det(e_{b}^{j})e_{i}^{a}$.

For $\beta = \i$, the action (\ref{eq:HolstAction}) 
can be written only with the left-handed variables:
\begin{equation}
 S_{\rm H}^{(+)} 
  = \frac{\i}{k}\int \Sigma^{(+)IJ} \wedge \Omega_{IJ}^{(+)} \ ,
  \label{eq:LeftHandedAction}
\end{equation}
where
\begin{equation}
 \Sigma^{(+)IJ} 
= \frac{1}{2}\left(e^{I} \wedge e^{J} 
- \frac{\i}{2} \epsilon^{IJ}{}_{KL}e^{K}\wedge e^{L}\right)\ ,
\end{equation}
and
\begin{equation}
 \Omega_{IJ}^{(+)} = 
  \frac{1}{2}\left( \Omega_{IJ} 
- \frac{\i}{2} \epsilon_{IJ}{}^{KL}\Omega_{KL} \right)\ .
\end{equation}
The sign $(+)$ explicitly denotes that the variable is left-handed, 
namely, $\beta = \i$.
The three constraints are
\begin{eqnarray}
 G_{i}^{(+)} 
  &=& (D_{a}^{(+)}E^{(+)a} )_{i} 
  = \partial_{a}E^{(+)}{}_{i}^{a} 
  + \epsilon_{ij}{}^{k}A^{(+)}{}_{a}^{j}E^{(+)}{}_{k}^{a} \ ,
  \\
 V_{b}^{(+)} 
  &=& E^{(+)}{}_{i}^{a}F^{(+)}{}_{ab}^{i} \ ,
  \\
 H^{(+)} 
  &=&
  -\frac{\i}{2\sqrt{ |\det(E^{(+)})}| }\epsilon^{ijk}
  E^{(+)}{}_{i}^{a}E^{(+)}{}_{j}^{b}F^{(+)}_{ab\, k} \ ,
\end{eqnarray}
where 
$E^{(+)}{}_{i}^{a} 
= \epsilon^{abc}\epsilon_{ijk}\Sigma^{(+)}{}_{bc}^{jk}$,
and $F^{(+)}{}_{ab}^{i}$ is the curvature of the connection 
$A^{(+)}{}_{a}^{i} = \Gamma_{a}^{i} + \i K_{a}^{i}$.
The wave function has to satisfy the quantized constraints, 
which are formally written as
\begin{equation}
 \hat{G}^{(+)}_{i}\Psi
 = \hat{V}^{(+)}_{b}\Psi 
 = \hat{H}^{(+)}\Psi 
 = 0 \ .
\end{equation}

In Ref. \cite{Mikovic2003}, 
it is suggested that the product of the group delta functions
\begin{equation}
 \Psi_{\rm BF}(A^{(+)})
  = \prod_{x \in \Sigma}\prod_{a,b}
  \delta\left( \e^{F^{(+)}_{ab}(x)} \right) 
  \label{eq:BFState}
\end{equation}
is a solution of the constraints. 
This state is originally derived from the formal integral 
$\int {\cal DB}\, 
\exp [\, \i S_{\rm BF}\, ]=\delta(\e^{F})$, 
where $S_{\rm BF}=\int_{\Sigma} \Tr (B \wedge F)$ 
is the $SU(2)$ BF action in 3D Euclidean space $\Sigma$.
Thus let us call state (\ref{eq:BFState}) the BF state.
The group delta function has the following properties:
\begin{eqnarray}
 \delta\left( g\, \e^{F^{(+)}_{ab}} g^{-1} \right) 
  &=& \delta \left( \e^{F^{(+)}_{ab}} \right) \ ,
  \label{eq:GaugeTransformationOfDelta}
  \\ 
  F^{(+)}_{ab}\delta \left( \e^{F^{(+)}_{ab}} \right) &=& 0 \ ,
\end{eqnarray}
where $g$ is an element of the gauge group.
Therefore the state $\Psi_{\rm BF}(A^{(+)})$ is gauge invariant and 
$\hat{V}^{(+)}_{b}\Psi_{\rm BF}(A^{(+)}) 
= \hat{H}^{(+)}\Psi_{\rm BF}(A^{(+)}) = 0$.
This state is proposed as a tool 
to construct a flat vacuum state \cite{Mikovic2004}.

\subsection{Chiral asymmetric extension}
Following the strategy of Ref. \cite{Randono2006}, we first consider
the chiral asymmetric model with purely imaginary values of $\beta$.
The left-handed action (\ref{eq:LeftHandedAction}) is extended to 
the chiral asymmetric one as follows:
\begin{eqnarray}
 S 
  &=&
  \alpha^{(+)}S_{\rm H}^{(+)} + \alpha^{(-)}S_{\rm H}^{(-)} 
  \nonumber \\
 &=&
  \frac{1}{4k}\int 
  \biggl[
  \left(\alpha^{(+)} + \alpha^{(-)}\right)
  \epsilon_{IJKL}\, e^{I} \wedge e^{J} \wedge \Omega^{KL}
  + 
  2\i \left(\alpha^{(+)} - \alpha^{(-)}\right)
  e^{I}\wedge e^{J} \wedge \Omega_{IJ}
  \biggr]  . \label{eq:ChiralAsymmetricAction}
\end{eqnarray}
Here $\alpha^{(+)}$ and $\alpha^{(-)}$ are mixing parameters 
of the left- and right-handed components.
The sign $(-)$ means right-handed, i.e., $\beta=-\i$.
To identify the action (\ref{eq:ChiralAsymmetricAction}) with 
(\ref{eq:HolstAction}), the following identities are
obtained:
\begin{equation}
 \alpha^{(+)} + \alpha^{(-)}=1\ , \hspace{5mm}
  \alpha^{(+)} - \alpha^{(-)} = \frac{\i}{\beta} \ .
\end{equation}
Note that, in the case of the left-handed action, 
these parameters take $\alpha^{(+)} = 1$ and $\alpha^{(-)}=0$. 
One can find that imaginary $\beta$ controls the degree of the chiral 
asymmetry.
In this model, the Poisson brackets of the canonical variables
$(A^{(+)}, E^{(+)})$ and $(A^{(-)}, E^{(-)})$ are
\begin{equation}
 \left\{
  A^{(\pm)}{}_{a}^{i}(x), E^{(\pm)}{}_{j}^{b}(y)
 \right\}
 =
 \pm \frac{\i k}{\alpha^{(\pm)}}\,
 \delta_{j}^{i}\delta_{a}^{b}\delta^{3}(x-y) \ . 
\end{equation}
Here $E^{(+)}{}_{i}^{a} = E^{(-)}{}_{i}^{a} = \det(e) e_{i}^{a}$ 
in the time gauge $e_{a}^{0}=0$;
nevertheless, these variables are treated independently of each other.
Each of the three constraints separates into 
left- and right-handed components independently.
Therefore, the extended wave function is given by
\begin{equation}
 \Psi(A^{(+)}, A^{(-)}) 
  =
  \prod_{x\in \Sigma}\prod_{a,b}
  \delta\left( \e^{\alpha^{(+)}F_{ab}^{(+)}(x)} \right)
  \delta\left( \e^{-\alpha^{(-)}F_{ab}^{(-)}{(x)}} \right) \ .
\end{equation}

\section{Generalized BF state}
 
\subsection{Real values of $\beta$}
To consider the extended BF state for generic real values of $\beta$,
new configuration variables are introduced:
\begin{eqnarray}
 A^{(-\frac{1}{\beta})}{}_{a}^{i} 
  &=&
  \alpha^{(+)}A^{(+)} + \alpha^{(-)}A^{(-)}
  = \Gamma_{a}^{i} - \frac{1}{\beta}K_{a}^{i} \ ,
  \\
 A^{(\beta)}{}_{a}^{i} 
  &=&
  \frac{1}{\alpha^{(+)}- \alpha^{(-)}}
  \left( \alpha^{(+)}A^{(+)} - \alpha^{(-)}A^{(-)} \right)
  = \Gamma_{a}^{i} + \beta K_{a}^{i} \ .
\end{eqnarray}
The corresponding momentum variables are
\begin{eqnarray}
 C_{i}^{a} 
  &=&
  \frac{1}{2\i}\left(E^{(+)}{}_{i}^{a} - E^{(-)}{}_{i}^{a}\right)
  = \epsilon^{abc}e_{b i}e_{c}^{0} \ ,
  \\
 E_{i}^{a} 
  &=& 
  \frac{1}{2}\left(E^{(+)}{}_{i}^{a} + E^{(-)}{}_{i}^{a}\right)
  = \det(e)e_{i}^{a}\ .
\end{eqnarray}
One can obtain the Poisson bracket relations as follows:
\begin{eqnarray}
 \left\{ A^{(-\frac{1}{\beta})}{}_{a}^{i}(x), C_{j}^{b}(y) \right\}
  &=&
  k \delta_{j}^{i}\delta_{a}^{b}\delta^{3}(x-y) \ ,
  \\
 \left\{ A^{(\beta)}{}_{a}^{i}(x), E_{j}^{b}(y) \right\}
  &=&
  k \beta \delta_{j}^{i}\delta_{a}^{b}\delta^{3}(x-y) \ .
\end{eqnarray} 
To construct the generalized BF state,
we attempt to define the extended BF (EBF) action
\begin{equation}
 S_{\rm EBF} = \int \Tr
  \left[
    \alpha^{(+)} e^{(+)} \wedge F^{(+)} 
    - \alpha^{(-)} e^{(-)} \wedge F^{(-)} 
  \right] \ ,
\end{equation}
where $e^{(\pm)}$ are triads playing the role of the $B$ field of 
the BF action and are written as
\begin{equation}
 e^{(\pm)}{}_{a}^{i} = e_{a}^{i} 
  = \frac{1}{2 \sqrt{|\det(E)|}}\epsilon_{abc}\epsilon^{ijk}
  E_{j}^{b}E_{k}^{c} \ .
\end{equation}
The action $S_{\rm EBF}$ is expressed in terms of the variables 
$A^{(-\frac{1}{\beta})}$ and $A^{(\beta)}$ as
\begin{eqnarray}
 S_{\rm EBF}
  &=&
  \frac{\i}{\beta}\int \Tr
  \left[
   e \wedge
   \left(
    F - \left(1 + \beta^{2} \right) K \wedge K
   \right)
  \right]
  \nonumber \\
 &=&
  \frac{\i}{\beta} \int \Tr
  \left[
   e \wedge
   \left(
    \left( 1 + \frac{1}{\beta^{2}}\right) R - \frac{1}{\beta^{2}}F 
    - \beta \d_{\Gamma}K
   \right)
  \right] \ , \label{eq:ExtendedBFActionWithRealBeta}
\end{eqnarray}
where $F$ and $R$ are the curvatures of the connections
$A^{(\beta)}$ and $\Gamma$ respectively,
and $\d_{\Gamma}K = \d K + [\Gamma, K]$.
The last term vanishes for the torsion-free condition
$\d_{\Gamma}e=0$.

One can propose an extended BF state defined in the following form:
\begin{eqnarray}
 \Psi(A^{(\beta)}, A^{(-\frac{1}{\beta})})
  &=&
  \int {\cal D}e\
  \exp
  \left[
   \frac{\i}{\alpha^{(+)} - \alpha^{(-)}}\ S_{\rm EBF}
  \right] 
  \nonumber \\
 &=& 
  \prod_{x \in \Sigma}\prod_{a,b}
  \delta
  \left(
   \exp \left[ \left(1 + \frac{1}{\beta^{2}}\right) R_{ab}(x) 
   - \frac{1}{\beta^{2}}F_{ab}(x) \right]
  \right) \ . \label{eq:TempGeneralizedBFState}
\end{eqnarray}
Note that when $\beta = \i$, 
state (\ref{eq:TempGeneralizedBFState}) keeps the
ordinary form (\ref{eq:BFState}).
This state has a problem.
Due to the gauge fixing $e_{a}^{0}=0$, 
the wave function should satisfy the additional constraint:
\begin{equation}
 \hat{C}_{i}^{a}\Psi 
  = - \i k \frac{\delta}{\delta A^{(-\frac{1}{\beta})}{}_{a}^{i}}\Psi
  = 0 \ .
\end{equation}
This equation implies that the wave function does not depend on the
variable $A^{(-\frac{1}{\beta})}$.
However, the connection $\Gamma$ included in the curvature $R$ 
is the explicit function of both $A^{(\beta)}$ and 
$A^{(-\frac{1}{\beta})}$, namely,
\begin{equation}
 \Gamma_{a}^{i} 
  = \frac{A^{(\beta)}{}_{a}^{i}
  +\beta^{2}A^{(-\frac{1}{\beta})}{}_{a}^{i}}{1+\beta^{2}}\ .
\end{equation}
To avoid this problem, we regard the connection $\Gamma$ 
as the explicit variable of $E$, instead of $A^{(\beta)}$ and 
$A^{(-\frac{1}{\beta})}$.
This can be done via the torsion-free condition $\d_{\Gamma}e=0$.
Taking this modification into account, the extended BF state
is redefined:
\begin{equation}
 \Psi_{R}(A^{(\beta)})
  = \prod_{x \in \Sigma}\prod_{a,b}
  \delta
  \left(
   \exp \left[ \left(1 + \frac{1}{\beta^{2}}\right) R_{ab}(x) 
   - \frac{1}{\beta^{2}}F_{ab}(x) \right]
  \right) \ . \label{eq:GeneralizedBFState}
\end{equation}
Although the state $\Psi_{R}(A^{(\beta)})$ has the same form as 
(\ref{eq:TempGeneralizedBFState}), it is 
the explicit function of $A^{(\beta)}$ only, 
and is parameterized by the Levi-Civita curvature $R$.
It is an analog of the fact that 
the ordinary wave function 
$\Psi_{p}(x) = \exp[\, -\i\, (Et - {\bf p}\cdot {\bf x}) \, ]$
can be regarded as the position function 
parameterized by the momentum.

\subsection{Constraints and inner products}
Here, we confirm whether state
(\ref{eq:GeneralizedBFState}) satisfies the three constraints with
real values of $\beta$.
The Gauss constraint requires the wave function to be
invariant under the $SU(2)$ gauge transformation.
The state $\Psi_{R}(A^{(\beta)})$ is gauge invariant
because of the property of the group delta function 
(\ref{eq:GaugeTransformationOfDelta}).

The simple inner product between two states can be supposed as
\begin{eqnarray}
 \langle \, \Psi_{R'} | \Psi_{R} \, \rangle
  &=&
  \int {\cal DA}\
  \Psi_{R'}^{\dagger}(A^{(\beta)})\Psi_{R}(A^{(\beta)})
  \nonumber \\
 &=&
  \prod_{x}\prod_{a, b}\delta
  \left(
   \exp \left[ \left(1 + \frac{1}{\beta^{2}}\right)
   \left( R_{ab} - R'_{ab} \right) \right]
  \right) 
  \nonumber \\
 &\equiv& \delta \left( R - R' \right) \ .
\end{eqnarray}
Here ${\cal DA}$ is the appropriate measure 
of the connection $A^{(\beta)}$ normalized such that 
$\int {\cal DA} = 1$.
This inner product is too sensitive.
When $R'$ takes a different value from $R$, it vanishes,
even if $R$ and $R'$ are in the equivalence class of $SU(2)$ gauge 
and diffeomorphism transformations.
To make the inner product more convenient, the following new 
inner product is introduced:
\begin{eqnarray}
 (\, \Psi_{R'}|\Psi_{R} \, )
  &=& \int {\cal D}\phi \
  \langle \, \Psi_{R'}|\ {\cal U}(\phi)\, | \Psi_{R} \, \rangle 
  \nonumber \\
 &=&
  \int {\cal D}\phi \ \delta
  \left(
   R - \phi  R' 
  \right) \ .
\end{eqnarray}
Here ${\cal U}(\phi)$ is an operator of the gauge and 
diffeomorphism transformations. 
The integral $\int {\cal D}\phi$ is over all of both 
transformations.
This construction of the inner product is an analogy of LQG
\cite{Perez2005}.
One can find that the dual state
\begin{equation}
 (\, \Psi_{R'} | 
  = \int {\cal D}\phi\ \langle \, \Psi_{R'}|\ {\cal U}(\phi) 
  = \int {\cal D}\phi\ \langle \, \Psi_{\phi R'} | 
\end{equation}
is a solution of the Gauss and diffeomorphism constraints.

Finally, we consider the Hamiltonian constraint
\begin{equation}
 H = - \frac{\beta}{2\sqrt{|\det(E)|}}
  \epsilon^{ijk}E_{i}^{a}E_{j}^{b}
  \left[
   F_{ab k} - \left(1 + \beta^{2} \right)
   \epsilon_{klm}K_{a}^{l}K_{b}^{m}
  \right] \ .
\end{equation}
Performing a similar calculation 
to (\ref{eq:ExtendedBFActionWithRealBeta}),
the smeared Hamiltonian constraint is deformed as
\begin{eqnarray}
 H(N) &=& \int \d^{3}x\ NH
  \nonumber \\
 &=&
  - \int \d^{3}x \ \frac{N\beta}{2\sqrt{|\det(E)|}}
  \epsilon^{ijk}E_{i}^{a}E_{j}^{b}
  \left[
  \left( 1 + \frac{1}{\beta^{2}} \right)R_{ab k}
  - \frac{1}{\beta^{2}}F_{ab k}
  \right] \ .
\end{eqnarray}
Therefore, if the Levi-Civita curvature operator $\hat{R}$ 
can be well defined, i.e., 
$\hat{R} \Psi_{R}(A^{(\beta)}) = R \Psi_{R}(A^{(\beta)})$,
then the state $\Psi_{R}(A^{(\beta)})$ will satisfy
\begin{equation}
 \int \d^{3}x\ \chi^{ab k} 
  \left[
   \left( 1 + \frac{1}{\beta^{2}} \right) \hat{R}_{ab k}
   - \frac{1}{\beta^{2}} \hat{F}_{ab k}
  \right]\ 
  \Psi_{R}(A^{(\beta)}) 
  = 0 \ ,
\end{equation} 
where $\chi$ is a test function.
According to Ref. \cite{Randono2006},
the Levi-Civita curvature operator $\hat{R}$ is defined as follows:
\begin{equation}
 \int \d^{3}x\ \chi^{ab k} \hat{R}_{ab k} 
  = \int {\cal D}\phi {\cal D}R'
  \left[
   \int \d^{3}x\ \chi^{ab k} \left(\phi R'_{ab k} \right)
  \right]
  | \Psi_{\phi R'} \, \rangle \langle \, \Psi_{\phi R'} | \ ,
\end{equation}
where the integral $\int {\cal D}R'$ is over the Levi-Civita 
curvature $R'$ modulo the equivalence class of the gauge and 
diffeomorphism transformations.
The action of this operator on the state $|\Psi_{R} \, \rangle$ 
becomes
\begin{eqnarray}
  \int \d^{3}x\ \chi^{ab k} \hat{R}_{ab k} |\Psi_{R} \, \rangle
  &=&
  \int {\cal D}\phi {\cal D}R' \
  \delta (R-\phi R')
  \left[
   \int \d^{3}x\ \chi^{ab k} \left(\phi R'_{ab k} \right)
  \right]
  | \Psi_{\phi R'} \, \rangle
  \nonumber \\
 &=&
  \int \d^{3}x\ \chi^{ab k}R_{ab k} |\Psi_{R}\, \rangle \ .
\end{eqnarray}
With this operator, one obtains 
\begin{equation}
 \hat{H}(N)\Psi_{R}(A^{(\beta)}) = 0 \ .
\end{equation}
Consequently, the state $\Psi_{R}(A^{(\beta)})$ satisfies 
all three constraints.

\section{Conclusions and discussion}
In this paper, we have constructed the generalized BF state for real
values of $\beta$.
This has been done via the chiral asymmetric extension. 
The generalized state is an explicit 
function of the connection $A^{(\beta)}$, and is parameterized by 
the Levi-Civita curvature $R$ as well as in the generalized CS state.
It is gauge invariant and solves all constraints
with the appropriate inner product and the operator.

This state would be associated with the space such that 
$(1+\beta^{2})R_{ab} - F_{ab} = 0$.
It contains a special case, i.e., a flat 3D space 
$R = F = 0$.
More discussions are necessary to obtain further specific 
interpretations.
Problems with the connection with generic $\beta$ may arise,
because this connection is not a pull-back of a space-time
connection \cite{Samuel2000}. 

It would be interesting to consider a loop representation 
of the state $\Psi_{R}(A^{(\beta)})$:
\begin{eqnarray}
 \Psi_{R}(\gamma) 
  &=& 
  \langle \, \gamma | \Psi_{R} \, \rangle
  = \int {\cal DA}\ \langle \,\gamma |A^{(\beta)} \, \rangle 
  \langle \, A^{(\beta)} |\Psi_{R}\, \rangle
  \nonumber \\
 &\sim&
  \int {\cal DA}\ W(A^{(\beta)}, \gamma) \prod_{x}\prod_{a,b}
  \delta 
  \left(\exp \left[ F_{ab} - (1+\beta^{2})R_{ab} \right] \right) \ .
  \label{eq:LoopRep1}
\end{eqnarray}
Here $W(A^{(\beta)}, \gamma)$ is a spin network with a graph 
$\gamma$, which is a generalized Wilson loop constructed 
from holonomy edges and invariant tensors.
The part
\begin{equation}
 \int {\cal DA}\ \prod_{x}\prod_{a,b}
  \delta 
  \left( \exp \left[F_{ab} - (1+\beta^{2})R_{ab} \right] \right)
\end{equation}
looks like a generating functional of the spin foam model with a 
source term, which is known as the Freidel--Krasnov (FK) model 
\cite{Freidel1999}.
Let us consider a discretized 3D space with a 
triangulation $\Delta$.
The corresponding dual cell $\Delta^{*}$ has vertices $v$,
edges $l$, and faces $f$. 
In the FK model, the discretized generating functional is given by 
\begin{eqnarray}
 Z_{\rm FK}[J]
 &=&
  \int {\cal DA DB} \ 
  \exp \left[ \i \int \Tr (B \wedge F + B \wedge J) \right]
  \nonumber \\
 &=&
  \int \prod_{l}\d g_{l}\ 
  \prod_{f} \sum_{\Lambda_{f} \in {\rm Irrep}} 
  \dim(\Lambda_{f})\,
   \Tr \left[ 
       R^{(\Lambda_{f})}  
       \left(
	g_{l_{1}} \e^{J_{v_{1}}} \cdots g_{l_{n}} \e^{J_{v_{n}}}
	\right)
      \right],
\end{eqnarray}
where $J$ is a source term for $B$, 
$R^{(\Lambda)}(g_{l})$
is a representation of the group element $g_{l}=\exp[\int_{l} A]$ 
with a spin label $\Lambda$, 
vertices $v_{1}, \cdots ,v_{n}$ and edges $l_{1}, \cdots ,l_{n}$ 
are associated with the $n$-polygonal $\partial f$,
and the sum is taken over all irreducible representations.
Thus the loop representation (\ref{eq:LoopRep1}) is expressed as 
\begin{eqnarray}
\Psi_{R}(\gamma)
  &=& \int \prod_{l}\d g_{l}\ W(A_{l}^{(\beta)}, \gamma)
  \prod_{f} \sum_{\Lambda_{f} \in {\rm Irrep}}
  \dim(\Lambda_{f})\,
  \Tr \left[ 
       R^{(\Lambda_{f})}
       \left(
	g_{l_{1}} \e^{r_{v_{1}}} \cdots g_{l_{n}} \e^{r_{v_{n}}}
       \right)
      \right],
\end{eqnarray}
where $r = -(1+\beta^{2})R$.
The limit $R \to 0$ is consistent with the spin network invariant
$\Psi_{R=0}(\gamma)$ in Ref. \cite{Mikovic2003}.

\section*{Acknowledgements}
The authors would like to thank 
H. Taira, T. Oka, and K. Eguchi for helpful discussions.

\end{document}